\documentclass[showpacs,twocolumn,superscriptaddress,prl,aps,longbibliography]{revtex4-1}

\usepackage{amsmath}
\usepackage{amssymb}
\usepackage{maybemath}
\usepackage{mathtools}
\usepackage{xcolor}
\usepackage{graphicx}
\usepackage{bm}
\usepackage{comment}
\usepackage[caption=false]{subfig}
\usepackage[export]{adjustbox}
\usepackage{braket}
\usepackage{mathrsfs}
\usepackage{tikz}

\DeclarePairedDelimiter\floor{\lfloor}{\rfloor}

\usetikzlibrary{decorations.pathmorphing}
\usetikzlibrary{shapes}
\usetikzlibrary{shapes.geometric}

\tikzset{
    magnetic/.style={
        fill,
        shape border rotate=-90,
        isosceles triangle,
        isosceles triangle apex angle=60,
        node distance=1,
        minimum height=.1
    }
}

\tikzset{
    othermagnetic/.style={
        fill,
        shape border rotate=90,
        isosceles triangle,
        isosceles triangle apex angle=60,
        node distance=1,
        minimum height=.1
    }
}

\usepackage{dcolumn}
\newcolumntype{.}{D{.}{.}{-1}}

\usepackage[%
  colorlinks=true,
  urlcolor=blue,
  linkcolor=blue,
  citecolor=blue
]{hyperref}

\makeatletter

\newcommand{\Rmnum}[1]{\expandafter\@slowromancap\romannumeral #1@}
\makeatother

\begin{document}

\title{The onset time of Fermi's golden rule}%

\newcommand{\addrMRS}{Aix Marseille Univ, CNRS, Centrale Marseille, Institut Fresnel, Marseille, France}
\newcommand{\addrSGP}{Current address: Institute of Materials Research and Engineering, A*STAR (Agency for Science Technology and Research), 138634 Singapore}
\newcommand{\addrMPIK}{Max Planck Institute for Nuclear Physics, Saupfercheckweg 1, 69117 Heidelberg, Germany}

\author{Vincent Debierre}
\email[]{vincent.debierre@mpi-hd.mpg.de}
\affiliation{\addrMPIK}

\author{Emmanuel Lassalle}
\affiliation{\addrMRS}
\affiliation{\addrSGP}

\begin{abstract}

Fermi's golden rule describes the decay dynamics of unstable quantum systems coupled to a reservoir, and predicts a linear decay in time. Although it arises at relatively short times, the Fermi regime does not take hold in the earliest stages of the quantum dynamics. The standard criterion in the literature for the onset time of the Fermi regime is $t_F\sim1/\Delta\omega$, with $\Delta\omega$ the frequency interval around the resonant transition frequency $\omega_0$ of the system, over which the coupling to the reservoir does not vary appreciably. In this work, this criterion is shown to be inappropriate in general for broadband reservoirs, where the reservoir coupling spectrum takes the form $R\left(\omega\right)\propto\omega^\eta$, and for which it is found that for $\eta>1$, the onset time of the Fermi regime is given by $t_F\propto\left(\omega_{\mathrm{X}}/\omega_0\right)^{\eta-1}\times1/\omega_0$ where $\omega_{\mathrm{X}}$ is the high-frequency cutoff of the reservoir. Therefore, the onset of the Fermi regime can take place at times orders of magnitude larger than those predicted by the standard criterion. This phenomenon is shown to be related to the excitation of the off-resonant frequencies of the reservoir at short times. For broadband reservoirs with $\eta\leq1$, and for narrowband reservoirs, it is shown that the standard criterion is correct. Our findings revisit the conditions of applicability of Fermi's golden rule and improve our understanding of the dynamics of unstable quantum systems. 

\end{abstract}


\maketitle

\textit{Introduction.---} Dirac's result \cite{DiracGolden} describing the decay of excited quantum systems as linear in time is so solid and ubiquitous as to having been
called \emph{the (second) golden rule} by Fermi \cite{FermiGolden,ItaliaFermi}. The result is widely used throughout physics, including in nuclear \cite{NuclFermiPhase,NuclFermiQuasi} and particle \cite{Weinberg1,PartFermiFrag} physics, atomic \cite{AtomFermiTwo,AtomFermiMirror}, molecular \cite{MolecFermiBenzene,MolecFermiDissoc} and plasma \cite{PlasmaFermiQuasi} physics, biological \cite{BioFermiPurple} and solid-state \cite{SolidFermiSurface,SolidFermiExciton,SolidFermiOrgCry} physics, as well as photonics \cite{PhotoFermiDirect}. It is valid at short to intermediate times, and naturally merges into the Weisskopf-Wigner exponential decay \cite{WW} at longer times. Deviations from the golden rule at very short and very long times are rather well understood. Arguments based on the Paley-Wiener theorem exclude the possibility of purely exponential decay at very long times \cite{Khalfin,FacchiPhD}. As far as very short times are concerned, the probability that the system survives in its initial excited state is known to decrease quadratically with time, a phenomenon known as the Zeno decay \cite{MisraSudarshan,KofKur,FaNaPa}. This quadratic decay can be derived by an expansion of the time-evolution operator up to the lowest nontrivial order in time. By the same token, the exit from the Zeno regime is understood to take place when the next (quartic) term in the same expansion becomes comparable to the quadratic term \cite{KarpovZeno}.

The onset of the linear decay described by Fermi's golden rule (FGR), however, is less readily understood. We first note that, although the Fermi regime is, in ascending order of time scales, the second clearly identified regime after the initial Zeno regime, the exit from the Zeno regime does not in general mark the onset of the Fermi regime. The approximately $t$-quartic decay that follows the initial Zeno dynamics has indeed little in common with Fermi's $t$-linear decay. The transition between the Zeno and Fermi regimes is clarified in the present work. We define the onset time $t_F$ of the golden rule as the time such that when $t\gg t_F$, the FGR is guaranteed to be valid. The golden rule is usually derived from time-dependent perturbation theory \cite{Bellac,Englert}, which assumes short times, but the derivation uses a mathematical property (a distributional limit) that only comes into play at sufficiently long times. There is no contradiction here as the characteristic times for perturbation theory and the distributional limit are different, but the question of what constitutes `sufficiently long times' for Fermi's linear regime to take over has seldom been examined in detail. We set out to bridge this gap in the present work. To do so, it will be necessary to distinguish between broadband reservoirs, which typically arise if the environment is open and has a continuous spectrum, and narrowband reservoirs, which typically arise if the environment is confined and has a discrete spectrum.

\textit{General equations.---} We consider a two-level system, consisting of a ground state $\ket{\mathrm{g}}$ and an excited state $\ket{\mathrm{e}}$
separated by the energy $\hbar\omega_0$, and initially prepared in $\ket{\mathrm{e}}$. Due to its coupling with the reservoir, the system will naturally decay to the ground state
$\ket{\mathrm{g}}$, with a survival probability $P_{\mathrm{surv}}\left(t\right)=1-\Gamma\left(t\right) t$. The generalised decay rate is given by first-order time-dependent
perturbation theory \cite{Englert,KofKur} (also see Supp. Mat.)
\begin{equation} \label{eq:General}
\Gamma\left(t\right)= 2\pi\int_0^{\infty}\mathrm{d}\omega\,F_t\left(\omega-\omega_0\right)R\left(\omega\right).
\end{equation}
Here the reservoir coupling spectrum (RSC) $R(\omega)$ is given by
\begin{equation}\label{eq:Reservoir}
R(\omega) = \hbar^{-2}\sum_k |\bra{\mathrm{e},0}\hat{H}_I\ket{\mathrm{g},1_{k}}|^2\delta(\omega-\omega_k)
\end{equation}
where the state of the reservoir $\ket{1_{k}}$ contains one excitation in the mode labelled by $k$ and $\ket{0}$ is the vacuum state of the reservoir. $\hat{H}_I$ is the system-reservoir interaction Hamiltonian. The function $F_t(\omega-\omega_0)$, on the other hand, corresponds to the spectral profile of the two-level system at time $t$:
\begin{equation} \label{eq:Spectral}
F_t\left(\omega-\omega_0\right) = \frac{t}{2\pi}\mathrm{sinc}^2\left((\omega-\omega_0)\frac{t}{2} \right).
\end{equation}
From Eq.~(\ref{eq:General}), the most straightforward derivation of the golden rule is as follows: it is seen that in the large time limit
$t\rightarrow+\infty$, $F_t\left(\omega-\omega_0\right)\rightarrow\delta\left(\omega-\omega_0\right)$ and Eq.~(\ref{eq:General}) gives: $\Gamma\left(t\right)\rightarrow2\pi R(\omega_0)\equiv\Gamma_0$, which is the natural decay rate given by the golden rule. Note that only the reservoir states of frequency $\omega_0$ contribute to the decay in the Fermi regime. This is to be contrasted with the Zeno regime, which appears in the limit of very short times $t\rightarrow 0$, and for which $\Gamma\left(t\right)\rightarrow A\times t$ with $A=\int_{0}^{+\infty}\mathrm{d}\omega\,R\left(\omega\right)$. The Zeno regime can be understood as that in which all the reservoir frequencies featured in the RSC respond in phase to the excitation from the two-level system \cite{EdouardIsa}.

The use of the distributional limit in the golden rule begs the question of when the large time limit is reached. When this happens depends on the shape of the RSC function $R(\omega)$, but this answer is unsatisfactorily vague.
In quantum mechanics textbooks \cite{Messiah2,Cohen2,Bellac,grynberg2010introduction,barnett2002methods} as well as in Ref.~\cite{ItaliaFermi},
the criterion $t\gg2\pi/\Delta\omega$ for the applicability of the golden rule is given, with $\Delta\omega$ the width of the interval around the transition frequency $\omega_0$ in which $R\left(\omega\right)\simeq R\left(\omega_0\right)$. Specific statements from the literature that illustrate our presentation of the standard criterion are given in the Supp. Mat. As we shall see, this criterion is not valid in the general case.

In order to evaluate the integral (\ref{eq:General}) in the most general case, it is tempting to try to expand the RSC function around the transition resonance frequency $\omega_0$. However, this yields a sum of divergent terms: another approach is needed.
We have found it convenient to distinguish between broadband and narrowband reservoirs. Broadband reservoirs feature all frequencies in their RSC (below a high-energy cutoff).
They can be treated by adequately modifying the naive series-expansion approach mentioned just above: the RSC function is the product of a function that will be Taylor-expanded around $\omega_0$, with a `cutoff function' that excludes very high frequencies from the dynamics. Narrowband reservoirs, on the other hand, only feature a small frequency window.
We will treat them here as having a RSC of the Breit-Wigner type. Neglecting threshold effects \cite{Threshold}, an approximation which we carefully justify, the dynamics can be solved for such reservoirs
through Laplace-Fourier transform techniques.

\textit{Broadband reservoirs.---} Let us start with the case of broadband reservoirs. They correspond in general to open environments with a continuous spectrum. Examples include the interaction between an atom and the electromagnetic modes in free space \cite{Seke,LassalleChampenois,lassalle:tel-02283698}, or that of a two-level system interacting with a sub- or super-Ohmic dissipative bath, a model that describes many different systems \cite{ChinaZenoCriterion,OhmBaths}, and in particular, atoms interacting with a degenerate Fermi gas \cite{TheoryTunnelInteract,LowTTheoryTunnelInteract}. In a sufficiently general setting, the RSC can be cast in the form
\begin{subequations} \label{eq:Broad}
\begin{equation} \label{eq:BroadRSC}
  R\left(\omega\right)=\lambda\,\omega\left(\frac{\omega}{\omega_{\mathrm{X}}}\right)^{\eta-1}F_{\mathrm{X}}\left(\omega\right)
\end{equation}
where the dimensionless parameter $\lambda\ll1$ measures the strength of the system-environment coupling and $\eta$ is in general a real, nonnegative parameter. Also, $F_{\mathrm{X}}$ is the cutoff function, that is close to unity up to $\omega\lesssim\omega_{\mathrm{X}}$ and close to zero from $\omega\gtrsim\omega_{\mathrm{X}}$ onwards \footnote{For instance, for atomic transitions, $F_{\mathrm{X}}\left(\omega\right)=\left[1+\left(\omega/\omega_{\mathrm{X}}\right)^2\right]^{-\mu}$ with $\mu\geq4$ an integer, and for sub- or super-Ohmic baths, $F_{\mathrm{X}}\left(\omega\right)=\exp\left(-\omega/\omega_{\mathrm{X}}\right)$.},
with $\omega_{\mathrm{X}}\gg\omega_0$ the cutoff frequency. We emphasise that the cutoff function is not introduced artificially:
in the case of electronic transitions in atoms, for instance, it appears from a rigorous calculation of the coupling between electrons and photons beyond the electric dipole
approximation \cite{Moses,Seke}. More general broadband reservoirs, for instance in the case of atomic transitions, can be written as a weighted sum of functions of the form
given in Eq.~(\ref{eq:BroadRSC}), with different powers $\eta$ \cite{Seke,LassalleChampenois,lassalle:tel-02283698}. From the golden rule, the natural decay rate for a reservoir of the form (\ref{eq:BroadRSC}) reads
\begin{equation} \label{eq:BroadRate}
  \Gamma_0\simeq2\pi\,\lambda\,\omega_0\left(\frac{\omega_0}{\omega_{\mathrm{X}}}\right)^{\eta-1}.
\end{equation}
\end{subequations}
In order to determine the onset of Fermi's linear decay regime for broadband reservoirs of the form (\ref{eq:BroadRSC}),
we derive (see details in Supp. Mat.) analytical expressions of the generalised decay rate (\ref{eq:General}) by distinguishing three time regimes:
(i) the cutoff regime $\omega_{\mathrm{X}}t\ll1$,
(ii) the intermediate regime $\omega_{0}t\ll1\ll\omega_{\mathrm{X}}t$,
and (iii) the resonant regime $\omega_0t\gg1$ \footnote{We call $\omega_0t\gg1$ the resonant regime because in this regime, the transition frequency $\omega_0$ of the two-level system has been spectrally resolved by the dynamics and starts playing its role as the resonance frequency.}. This will allow for different approximations of the frequency integral in Eq.~(\ref{eq:General}).

The case where $\eta$ is a strictly positive integer in (\ref{eq:BroadRSC}) has been investigated in detail in Refs.~\cite{LassalleChampenois,lassalle:tel-02283698} in the resonant  regime $\omega_0t\gg1$, and
we extend here the results to arbitrary positive values of $\eta$.
In order to do so, we first write $\omega^\eta=\omega^{\floor{\eta}}\omega^{\eta-\floor{\eta}}$
in the integral $\Gamma\left(t\right)$ [Eq.~(\ref{eq:General})], where $\floor{\eta}$ denotes the integer part of  $\eta$, and expand $\omega^{\floor{\eta}}$
around the transition frequency $\omega_0$ at all orders. Moreover, for the calculations, we do not specify the form of $F_{\mathrm{X}}(\omega)$ in Eq.~(\ref{eq:BroadRSC}) (this necessitates an approximate computation of the generalised decay rate (\ref{eq:General})), and we will use the notation $\int_0^{+\infty}\mathrm{d}\omega\,F_{\mathrm{X}}\left(\omega\right)\equiv C\omega_{\mathrm{X}}$ where $C$ is typically of the order of unity. The decay rate then features two contributions: a resonant contribution, and a tail contribution.

The first, resonant contribution $\Gamma^{\mathrm{res}}\left(t\right)$ is due to the constant term in the binomial expansion of $\omega^{\floor{\eta}}$.
For this term, only the part of $F_t\left(\omega-\omega_0\right)$ that probes the reservoir $R\left(\omega\right)$ in Eq.~(\ref{eq:General}) in a frequency range of width $2\pi/t$ around $\omega_0$ contributes to the decay. We make the approximation \cite{LassalleChampenois} that $F_t\left(\omega-\omega_0\right)=t/\left(2\pi\right)$ in the interval $-\pi/t<\omega-\omega_0<\pi/t$ and vanishes elsewhere. As mentioned above, we need to distinguish three time regimes (see Supp. Mat.): (i) In the cutoff regime $\omega_{\mathrm{X}}t\ll1$, (our approximation of) $F_t$ probes the entire RSC, namely, the frequency range between $0$ and $\omega_{\mathrm{X}}$. This yields
\begin{subequations} \label{eq:BroadRes}
\begin{equation} \label{eq:BroadResCut}
  \Gamma_{\mathrm{cut}}^{\mathrm{res}}\left(t\right)\simeq\frac{1}{2\pi}\,\frac{C}{\eta-\floor{\eta}+1}\left(\frac{\omega_{\mathrm{X}}}{\omega_0}\right)^{\eta-\floor{\eta}+1}\left(\omega_0t\right)\Gamma_0.
\end{equation}
(ii) In the intermediate regime $\omega_{0}t\ll1\ll\omega_{\mathrm{X}}t$, $F_t$ probes the frequency range between $0$ and $\omega_0+\pi/t$, to give
\begin{equation} \label{eq:BroadResInt}
  \Gamma_{\mathrm{int}}^{\mathrm{res}}\left(t\right)\simeq\frac{1}{2\pi}\,\frac{1}{\eta-\floor{\eta}+1}\left(1+\frac{\pi}{\omega_0 t}\right)^{\eta-\floor{\eta}+1}\left(\omega_0t\right)\Gamma_0.
\end{equation}
(iii) In the resonant regime $\omega_0t\gg1$, studied in detail in Refs.~\cite{LassalleChampenois,lassalle:tel-02283698} in the framework of emission under repeated measurements (the calculations are identical to those of the present free-dynamics case), $F_t$ probes the frequency range between $\omega_0-\pi/t$ and $\omega_0+\pi/t$. The result in this regime reads
\begin{equation} \label{eq:BroadResRes}
  \Gamma_{\mathrm{res}}^{\mathrm{res}}\left(t\right)\simeq\Gamma_0.
\end{equation}
\end{subequations}

The second, tail contribution $\Gamma^{\mathrm{tail}}\left(t\right)$, which \emph{only exists for reservoirs for which $\eta\geq1$}, comes from all the other terms $k\geq1$
in the binomial expansion: this term includes a contribution from the \emph{off-resonant} modes of the reservoir to the decay. This is a crucial point as off-resonant frequencies are often overlooked, and we will see that these modes \emph{delay} the establishement of the FGR regime compared to the standard criterion $1\sim\Delta\omega\,t_F$. Again, for our calculations, we distinguish the same time regimes (see Supp. Mat.): (i) In the cutoff regime $\omega_{\mathrm{X}}t\ll1$, the argument of the square cardinal sine in $F_t$ (\ref{eq:Spectral}) is much smaller than $1$ for the whole frequency range of the reservoir, so that, Taylor-expanding the sine at the lowest order, we obtain
\begin{subequations} \label{eq:BroadTail}
\begin{equation} \label{eq:BroadTailCut}
  \Gamma_{\mathrm{cut}}^{\mathrm{tail}}\left(t\right)\simeq\frac{1}{2\pi}\frac{C}{\eta+1}\left(\frac{\omega_{\mathrm{X}}}{\omega_0}\right)^{\eta+1}\left(\omega_0 t\right)\Gamma_0,\qquad\eta\geq1.
\end{equation}
(ii) In the intermediate regime $\omega_0 t\ll1\ll\omega_{\mathrm{X}}t$, we make the same approximation as just above in the range $0\leq\omega\leq2\omega_0$, and for $2\omega_0\leq\omega$, we approximate \cite{LassalleChampenois,Messiah2} the square sine by its mean value $1/2$. The contribution from the latter interval is by far dominant, and we get
\begin{equation} \label{eq:BroadTailInt}
  \Gamma_{\mathrm{int}}^{\mathrm{tail}}\left(t\right)\simeq\frac{1}{\pi}\frac{C}{\eta-1}\left(\frac{\omega_{\mathrm{X}}}{\omega_0}\right)^{\eta-1}\frac{1}{\omega_0 t}\Gamma_0,\qquad\eta>1.
\end{equation}
(iii) In the resonant regime $\omega_0t\gg1$, we approximate the square sine by its mean value $1/2$ for all $\omega$ and we obtain the same result as in the intermediate regime:
\begin{equation} \label{eq:BroadTailRes}
  \Gamma_{\mathrm{res}}^{\mathrm{tail}}\left(t\right)\simeq\frac{1}{\pi}\frac{C}{\eta-1}\left(\frac{\omega_{\mathrm{X}}}{\omega_0}\right)^{\eta-1}\frac{1}{\omega_0 t}\Gamma_0,\qquad\eta>1.
\end{equation}
\end{subequations}
It is immediately seen that the results~(\ref{eq:BroadTailInt}) and (\ref{eq:BroadTailRes}) do not apply to the special case $\eta=1$. In this case, we have instead
\begin{equation} \label{eq:EtaOne}
  \Gamma_{\mathrm{int}}^{\mathrm{tail}}\left(t\right)\simeq\Gamma_{\mathrm{res}}^{\mathrm{tail}}\left(t\right)\simeq\frac{1}{\pi}C\log\left(\frac{\omega_{\mathrm{X}}}{\omega_0}\right)\frac{1}{\omega_0 t}\Gamma_0.
\end{equation}
We show in the Supp. Mat. that the approximations made to obtain analytical expressions of the decay rate
$\Gamma\left(t\right)\simeq\Gamma^{\mathrm{res}}\left(t\right)+\Gamma^{\mathrm{tail}}\left(t\right)$
are satisfactory for the three different time regimes, by comparing our results with the numerical computation of Eq.~(\ref{eq:General}).
\begin{figure}[t!]
  \includegraphics[height=.55\linewidth,valign=c]{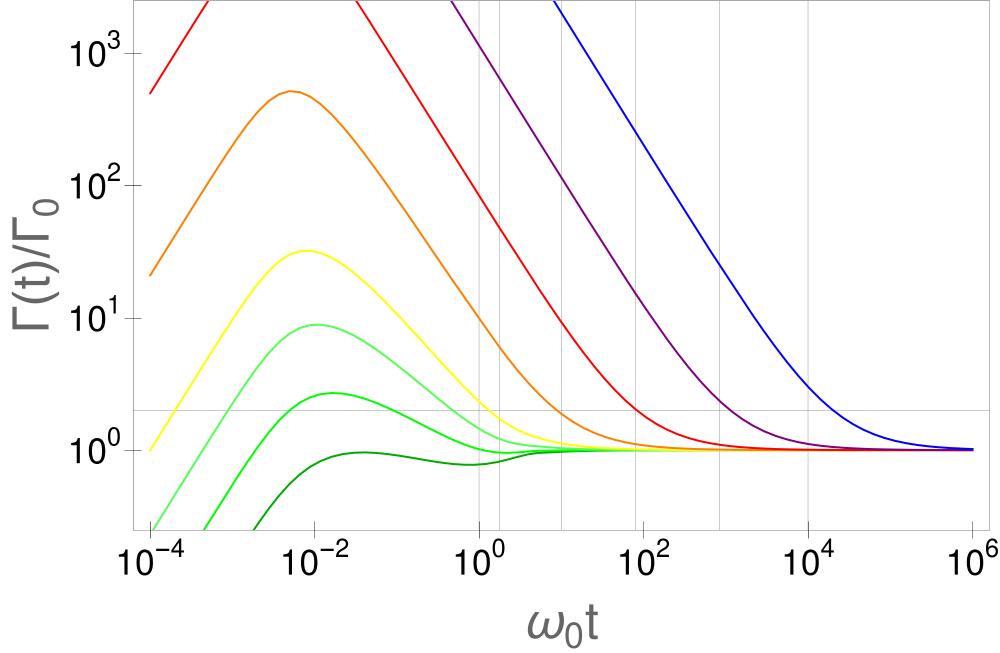}%
  \hfill
  \includegraphics[height=.375\linewidth,valign=c]{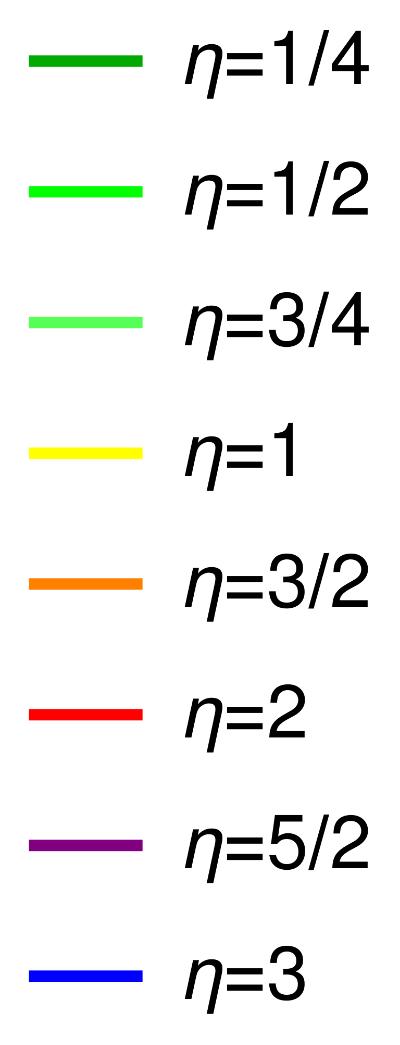}%
\caption{Ratio $\Gamma\left(t\right)/\Gamma_0$ of the generalised decay rate at time $t$ [Eq.~(\ref{eq:General})] and the decay rate given by the FGR (\ref{eq:BroadRate}) for broadband reservoirs (\ref{eq:BroadRSC}), as a function of $t$, computed numerically for various values of $\eta$. We have taken $F_{\mathrm{X}}\left(\omega\right)=\exp\left(-\omega/\omega_{\mathrm{X}}\right)$ and $\omega_{\mathrm{X}}=250\omega_0$. The vertical lines represent $t_F$ as determined in the text, for the various values of $\eta$ [see Eq.~(\ref{eq:BroadFermi})]. The horizontal line corresponds to $\Gamma\left(t_F\right)/\Gamma_0\simeq2$ in the cases for which $\eta>1$. \label{fig:BroadAll}}
\end{figure}

From these analytical expressions, we now aim at determining the onset time $t_F$ of the golden rule. We first note from Eqs.~(\ref{eq:BroadResCut}) and (\ref{eq:BroadTailCut}) that in the cutoff regime $\omega_{\mathrm{X}}t\ll1$, the decay rate $\Gamma_{\mathrm{cut}}^{\mathrm{res}}\left(t\right)+\Gamma_{\mathrm{cut}}^{\mathrm{tail}}\left(t\right)$ is proportional to $t$, yielding the quadratic decay characteristic of the Zeno regime: $P_{\mathrm{decay}}\left(t\right)\equiv 1-P_{\mathrm{surv}}\left(t\right)\propto t^2$. While our treatment does not give detailed information about when the dynamics exits this Zeno regime, it shows that in the intermediate regime $\omega_0t\ll1\ll\omega_{\mathrm{X}}t$, the Zeno dynamics no longer holds, as seen from Eqs.~(\ref{eq:BroadResInt}) and (\ref{eq:BroadTailInt}).

The regime in which the golden rule is established depends on the exponent $\eta$, and we must distinguish two cases. For the reservoirs with $\eta<1$, where there are no tail contributions to the decay rate, we can see from Eqs.~(\ref{eq:BroadResInt}) and (\ref{eq:BroadResRes}) that the golden rule is already established in the resonant regime $\omega_0t\gg1$, while it is not yet in the intermediate regime $\omega_0t\ll1\ll\omega_{\mathrm{X}}t$. Hence we conclude that the golden rule establishes \emph{between} these two regimes, and we set the onset time $t_F$ of the golden rule to be
\begin{subequations} \label{eq:BroadFermi}
\begin{equation} \label{eq:BroadFermiSmaller}
  t_F=\frac{1}{\omega_0},\quad\eta<1.
\end{equation}
For the reservoirs with $\eta\geq1$, where there are tail contributions to the decay rate, the golden rule is established later, in the resonant regime $\omega_0t\gg1$, as seen from Eqs.~(\ref{eq:BroadResRes}), (\ref{eq:BroadTailRes}) and (\ref{eq:EtaOne}). Explicitly, we see from these equations 
that the time $t_F$ at which the Fermi regime takes over reads
\begin{align}
  t_F&=\frac{1}{\pi}\frac{C}{\eta-1}\left(\frac{\omega_{\mathrm{X}}}{\omega_0}\right)^{\eta-1}\frac{1}{\omega_0},\quad&\eta>1, \label{eq:BroadFermiLa}\\
  t_F&=\frac{1}{\pi}\,C\log\left(\frac{\omega_{\mathrm{X}}}{\omega_0}\right)\frac{1}{\omega_0},\quad&\eta=1 \label{eq:BroadFermiEq}
\end{align}
where $C/\left(\eta-1\right)$ and $C\log\left(\omega_{\mathrm{X}}/\omega_0\right)$ are numerical factors broadly of the order of unity.
The case $\eta=1$ is somewhat specific mathematically, but it is very important physically as it describes atomic electronic transitions of the electric dipole
type \cite{Seke,LassalleChampenois} as well as the case of Ohmic disipative reservoirs \cite{OhmBaths}, the quantum equivalent of mechanical systems submitted to a
friction force proportional to their velocity. 
\end{subequations}
Note that Eq.~(\ref{eq:BroadFermiSmaller}), if extended to $\eta=1$, agrees with the more specific result Eq.~(\ref{eq:BroadFermiEq}), up to a numerical factor broadly of the order of unity.

There are thus two clearly different cases for the onset of the FGR: as $\eta$ is increased from $0$ to $1$, the onset of the golden rule always takes place around $t_F=1/\omega_0$, but as soon as $\eta>1$, the scaling changes: $t_F$ increases by the factor $\left(\omega_{\mathrm{X}}/\omega_0\right)^{\eta-1}$, which can be very large since in general $\omega_{\mathrm{X}}\gg\omega_0$ \cite{Moses,Seke}. This can be visualised in Fig.~\ref{fig:BroadAll} \footnote{Note also from Fig.~\ref{fig:BroadAll} that for $\eta\ll1$, the generalised decay rate verifies $\Gamma\left(t\right)\simeq\Gamma_0$ long before $t\sim1/\omega_0$ but however, $\Gamma\left(t\right)$ is not constant even at the lowest order of approximation in this regime, which precludes from including it in the Fermi regime.}.

Our result is in contradiction with the standard criterion \cite{Messiah2,Cohen2,Bellac,ItaliaFermi}, which has apparently remained largely unchallenged in the literature. Indeed, as described above, the standard criterion indicates that $t_F\simeq 1/\Delta\omega$ with $\Delta\omega$ the width of the interval around $\omega_0$ in which $R\left(\omega\right)\simeq R\left(\omega_0\right)$. But since $R\left(\omega_0\pm\delta\omega\right)\simeq R\left(\omega_0\right)\left(1\pm\eta\times\delta\omega/\omega_0\right)$ for broadband reservoirs, we can see that $\Delta\omega\simeq\omega_0/\eta$ and hence that if this criterion were true, then we would have $t_F\simeq\eta/\omega_0$, that is, $t_F\sim1/\omega_0$ for all broadband reservoirs regardless of the exponent $\eta$, to be contrasted with our results [Eq.~(\ref{eq:BroadFermi})]. Only for $\eta\leq1$, does our treatment vindicates the standard criterion.

\textit{Narrowband reservoirs.---}  As an archetype of a narrowband reservoir, we consider the case of a Breit-Wigner (also known as Cauchy-Lorentz) reservoir, for which the RSC reads:
\begin{equation}
\label{eq:lorentz}
R(\omega)=\frac{\kappa}{\pi}\frac{g^2}{(\omega-\omega_c)^2+\kappa^2}.
\end{equation}
This type of reservoir is encountered when considering the coupling of a quantum emitter with a single resonance, that can be a cavity, a photonic or a plasmonic resonance \cite{raimond2006exploring,ching1987dielectric,dung2001decay,delga2014quantum,varguet2016dressed}. The resonance is characterized by the frequency $\omega_c$ and width $\kappa$, and the coupling of the atom with this resonance by the coupling strength $g$.

\begin{figure}[t!]
  \centering
  \includegraphics[width=.775\linewidth,valign=c]{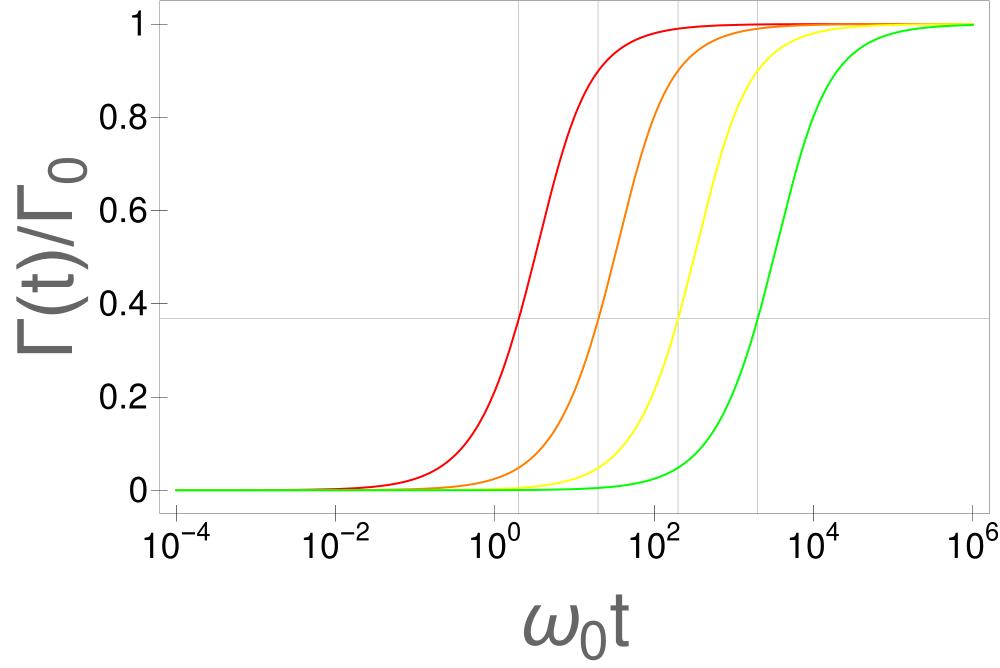}
  \includegraphics[width=.2\linewidth,valign=c]{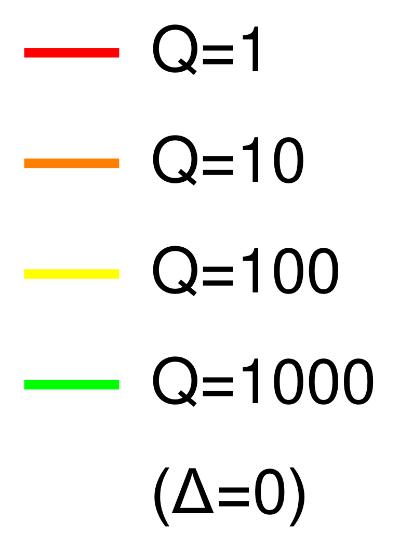}
    \caption{Ratio $\Gamma\left(t\right)/\Gamma_0$ as a function of $t$, for narrowband reservoirs
    (\ref{eq:gamma_narrowband_res}) in the
    resonant case $\omega_0=\omega_c$, with different quality factors
    $Q\equiv \omega_c/(2\kappa)$: $Q=1$ (red), $Q=10$ (orange),
    $Q=100$ (yellow) and $Q=1000$ (green). The horizontal line
    corresponds to $\Gamma\left(t\right)=\Gamma_0/\mathrm{e}$, and the vertical lines show
    the different times $t_F=2Q/\omega_c$. }
  \label{fig:quality_factor}
\end{figure}

In order to derive an analytical expression of the generalized decay rate $\Gamma\left(t\right)$, we make the approximation of extending the integral over the (nonexistent) negative frequencies of the reservoir spectrum in Eq.~(\ref{eq:General}). The validity of this approximation is
tested numerically in the Supp. Mat. and shown to work for any realistic plasmonic or photonic reservoir.
Considering first the resonant case where the resonance frequencies of the system and the reservoir are matched ($\omega_0=\omega_c$), we obtain (see Supp. Mat.):
\begin{equation}
\frac{\Gamma\left(t\right)}{\Gamma_0}=1-\frac{1-\mathrm{e}^{-\kappa t}}{\kappa t}
\label{eq:gamma_narrowband_res}
\end{equation}
where $\Gamma_0=2g^2/\kappa$. From this expression, it is natural to define the time $t_F$ at which the Fermi regime takes over, as the characteristic time of the decaying exponential. We thus take $t_F\equiv1/\kappa$, which is in agreement with the standard criterion \cite{Messiah2,Cohen2,Bellac,grynberg2010introduction,barnett2002methods,ItaliaFermi}.
Introducing the quality factor $Q\equiv \omega_c/(2\kappa)$ of the resonance, this criterion can be written as
\begin{equation}
t_F=2\frac{Q}{\omega_c}\, ,
\label{eq:tf_narrow}
\end{equation}
which shows that $t_F$ is independent of the coupling strength $g$, and scales linearly with the quality factor $Q$.
Therefore, from Eq.~(\ref{eq:tf_narrow}), we can conclude that for narrowband reservoirs and in the resonant case, \emph{losses} (that is, a wide spectrum) will \emph{accelerate}
the onset of the Fermi regime.
This is illustrated in Fig.~\ref{fig:quality_factor}, where we show how the decay rate $\Gamma\left(t\right)$ calculated from
Eq.~(\ref{eq:gamma_narrowband_res}) converges to the golden rule value $\Gamma_0$, for different quality factors $Q$. For a typical plasmonic resonance, $Q\sim 10$ and for the optical frequency $\omega_c=350$ THz (see \emph{e.g.} \cite{aljunid2016atomic}), $t_F=5.71\times 10^{-14}$ s, so the Fermi golden rule is established almost immediately. On the other hand, dielectric optical cavities, such as whispering gallery cavities, can have ultrahigh $Q$-factors with values over $10^8$ (see Ref.~\cite{jiang2017ultra} and Refs. therein). For such a high $Q$-factor, and assuming the same central frequency, the onset time of the golden rule is a few tens of microseconds, which cannot be considered as immediate.

Note that for very short times, we get from Eq.~(\ref{eq:gamma_narrowband_res}): $\Gamma\left(t\right)\simeq \kappa t \Gamma_0/2$,
which leads to a quadratic decay for the survival probability $P_\mathrm{surv}(t)$ characteristic of the Zeno regime.

\begin{figure}[t!]
  \centering
  \includegraphics[width=.775\linewidth,valign=c]{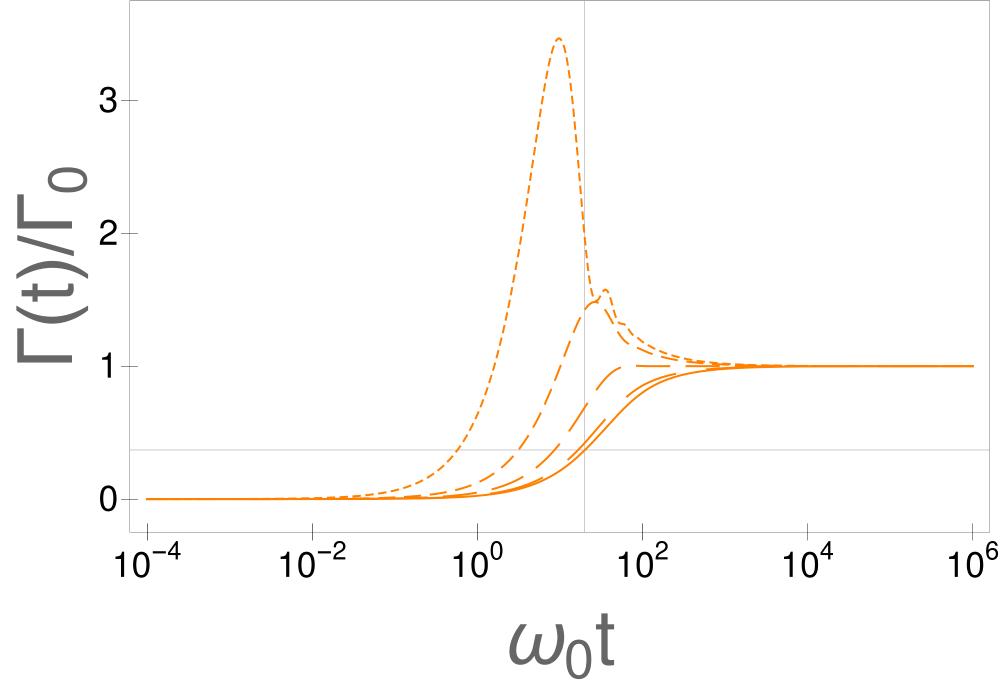}
  \includegraphics[width=.2\linewidth,valign=c]{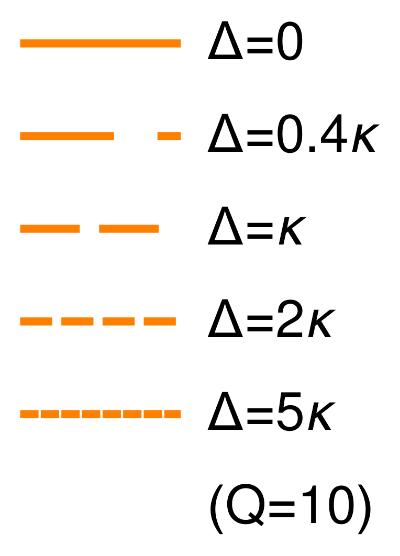}
  \caption{Ratio $\Gamma\left(t\right)/\Gamma_0$ as a function of $t$, for narrowband reservoirs
    (\ref{eq:gamma_narrowband_res}), and for different frequency
    detunings $\Delta=\omega_0-\omega_c$: $\Delta=0$ (solid),
    $\Delta=0.4\kappa$ (sparsely dashed), $\Delta=\kappa$ (dashed),
    $\Delta=2\kappa$ (densely dashed) and $\Delta=5\kappa$ (dotted). The quality factor is $Q=10$
    in all cases.}
  \label{fig:detuning}
\end{figure}

This criterion remains valid in the presence of detuning: $\Delta\equiv\omega_0-\omega_c\neq 0$. Indeed, in this case, we obtain (see Supp. Mat.)
\begin{equation}
\frac{\Gamma\left(t\right)}{\Gamma_0}=1-V\left[\frac{1-\mathrm{cos}\left(\Delta t\right)\mathrm{e}^{-\kappa t}}{\kappa t}\right]-\left(1-V\right)\mathrm{sinc}\left(\Delta t\right)\mathrm{e}^{-\kappa t}
\label{eq:gamma_narrowband_nonres}
\end{equation}
where $V\equiv[1-(\Delta/\kappa)^2]/[1+(\Delta/\kappa)^2]$ is equal to $1$ in the resonant case [and in this case
  Eq.~(\ref{eq:gamma_narrowband_nonres}) reduces to Eq.~(\ref{eq:gamma_narrowband_res})]. In Fig.~\ref{fig:detuning}, we show the ratio
$\Gamma\left(t\right)/\Gamma_0$ calculated from Eq.~(\ref{eq:gamma_narrowband_nonres}) for different detunings $\Delta$, and we see that the
time $t_F$ at which the Fermi regime is established does not change in the presence of detuning. Around the transition frequency, the RSC can be expanded as $R\left(\omega_0\pm\delta\omega\right)\simeq R\left(\omega_0\right)\left[1\pm2\left(\delta\omega/\kappa\right)^2\right]$, which means that $\Delta\omega=\kappa/2$ and hence, that the onset time of the linear decay regime predicted by the standard criterion \cite{Messiah2,Cohen2,Bellac,ItaliaFermi} is confirmed by our result Eq.~(\ref{eq:tf_narrow}). However, we notice (also see Ref.~\cite{ChinaZenoCriterion}) different behaviors before the onset of the golden rule: as long as $\Delta\leq\kappa$, the generalized decay rate is always smaller than $\Gamma_0$ (Zeno effect), whereas for $\Delta>\kappa$, the decay rate is greatly enhanced during a short period of time (anti-Zeno effect).

\textit{Discussion and conclusion.---} The time $t_F$ at which Fermi's golden rule starts being a good approximation to the decay dynamics of a two-level system was obtained.
It was found that, in the case of broadband reservoirs with $\eta\geq1$, $t_F\propto\left(\omega_{\mathrm{X}}/\omega_0\right)^{\eta-1}\times1/\omega_0$, in stark disagreement with the standard criterion given in the literature \cite{Messiah2,Cohen2,Bellac,ItaliaFermi}, which gives $t_F\sim1/\omega_0$ while for $\eta\leq1$, $t_F=1/\omega_0$, in agreement with the standard criterion (for $\eta=1$ we can use either result, as they agree up to numerical prefactors of the order of unity). 

In the case of narrowband reservoirs (here modeled by a Breit-Wigner coupling spectrum), we have obtained the universal result $t_F=2Q/\omega_c=1/\kappa$, where $Q$, $\omega_c$ and $\kappa$ are respectively the quality factor, central frequency and width of the reservoir. As can be seen, a high quality factor causes a late onset of the golden rule regime. Our analysis thus shows that studying the behaviour of the reservoir coupling spectrum around the transition frequency is not sufficient in the general case to determine the onset time of Fermi's golden rule regime, and emphasises the importance \cite{EdouardIsa} of the oft-overlooked off-resonant frequencies of the reservoir. Only for the slowly-growing broadband reservoirs $\eta\leq1$, and for narrowband reservoirs, for which off-resonant frequencies are a less dominant feature of the reservoir coupling spectrum, did we confirm the standard analysis of the onset of the Fermi regime.

We acknowledge helpful conversations with Thomas Durt and Caroline Champenois.

\bibliography{../../TwoLoop/Biblio}

\begin{thebibliography}{50}%
\makeatletter
\providecommand \@ifxundefined [1]{%
 \@ifx{#1\undefined}
}%
\providecommand \@ifnum [1]{%
 \ifnum #1\expandafter \@firstoftwo
 \else \expandafter \@secondoftwo
 \fi
}%
\providecommand \@ifx [1]{%
 \ifx #1\expandafter \@firstoftwo
 \else \expandafter \@secondoftwo
 \fi
}%
\providecommand \natexlab [1]{#1}%
\providecommand \enquote  [1]{``#1''}%
\providecommand \bibnamefont  [1]{#1}%
\providecommand \bibfnamefont [1]{#1}%
\providecommand \citenamefont [1]{#1}%
\providecommand \href@noop [0]{\@secondoftwo}%
\providecommand \href [0]{\begingroup \@sanitize@url \@href}%
\providecommand \@href[1]{\@@startlink{#1}\@@href}%
\providecommand \@@href[1]{\endgroup#1\@@endlink}%
\providecommand \@sanitize@url [0]{\catcode `\\12\catcode `\$12\catcode
  `\&12\catcode `\#12\catcode `\^12\catcode `\_12\catcode `\%12\relax}%
\providecommand \@@startlink[1]{}%
\providecommand \@@endlink[0]{}%
\providecommand \url  [0]{\begingroup\@sanitize@url \@url }%
\providecommand \@url [1]{\endgroup\@href {#1}{\urlprefix }}%
\providecommand \urlprefix  [0]{URL }%
\providecommand \Eprint [0]{\href }%
\providecommand \doibase [0]{http://dx.doi.org/}%
\providecommand \selectlanguage [0]{\@gobble}%
\providecommand \bibinfo  [0]{\@secondoftwo}%
\providecommand \bibfield  [0]{\@secondoftwo}%
\providecommand \translation [1]{[#1]}%
\providecommand \BibitemOpen [0]{}%
\providecommand \bibitemStop [0]{}%
\providecommand \bibitemNoStop [0]{.\EOS\space}%
\providecommand \EOS [0]{\spacefactor3000\relax}%
\providecommand \BibitemShut  [1]{\csname bibitem#1\endcsname}%
\let\auto@bib@innerbib\@empty
\bibitem [{\citenamefont {Dirac}(1927)}]{DiracGolden}%
  \BibitemOpen
  \bibfield  {author} {\bibinfo {author} {\bibfnamefont {P.~A.~M.}\
  \bibnamefont {Dirac}},\ }\bibfield  {title} {\enquote {\bibinfo {title} {The
  quantum theory of the emission and absorption of radiation},}\ }\href
  {http://rspa.royalsocietypublishing.org/content/114/767/243} {\bibfield
  {journal} {\bibinfo  {journal} {Proc. Royal Soc. A}\ }\textbf {\bibinfo
  {volume} {114}},\ \bibinfo {pages} {243} (\bibinfo {year}
  {1927})}\BibitemShut {NoStop}%
\bibitem [{\citenamefont {Fermi}(1950)}]{FermiGolden}%
  \BibitemOpen
  \bibfield  {author} {\bibinfo {author} {\bibfnamefont {E.}~\bibnamefont
  {Fermi}},\ }\href@noop {} {\emph {\bibinfo {title} {Nuclear Physics}}}\
  (\bibinfo  {publisher} {University of Chicago Press},\ \bibinfo {year}
  {1950})\BibitemShut {NoStop}%
\bibitem [{\citenamefont {Facchi}\ and\ \citenamefont
  {Pascazio}(1999)}]{ItaliaFermi}%
  \BibitemOpen
  \bibfield  {author} {\bibinfo {author} {\bibfnamefont {P.}~\bibnamefont
  {Facchi}}\ and\ \bibinfo {author} {\bibfnamefont {S.}~\bibnamefont
  {Pascazio}},\ }\href@noop {} {\emph {\bibinfo {title} {La regola d'oro di
  Fermi}}}\ (\bibinfo  {publisher} {Bibliopolis},\ \bibinfo {year}
  {1999})\BibitemShut {NoStop}%
\bibitem [{\citenamefont {Pai{\'c}}\ and\ \citenamefont
  {Antolkovi{\'c}}(1981)}]{NuclFermiPhase}%
  \BibitemOpen
  \bibfield  {author} {\bibinfo {author} {\bibfnamefont {G.}~\bibnamefont
  {Pai{\'c}}}\ and\ \bibinfo {author} {\bibfnamefont {B.}~\bibnamefont
  {Antolkovi{\'c}}},\ }\bibfield  {title} {\enquote {\bibinfo {title} {Effect
  of the phase space factor in the breakup of composite particles},}\ }\href
  {https://journals.aps.org/prc/abstract/10.1103/PhysRevC.23.1839} {\bibfield
  {journal} {\bibinfo  {journal} {Phys. Rev. C}\ }\textbf {\bibinfo {volume}
  {23}},\ \bibinfo {pages} {1839} (\bibinfo {year} {1981})}\BibitemShut
  {NoStop}%
\bibitem [{\citenamefont {Guttormsen}\ \emph {et~al.}(2011)\citenamefont
  {Guttormsen}, \citenamefont {Larsen}, \citenamefont {B{\"u}rger},
  \citenamefont {G{\"o}rgen}, \citenamefont {Harissopulos}, \citenamefont
  {Kmiecik}, \citenamefont {Konstantinopoulos}, \citenamefont {Krti{\v c}ka},
  \citenamefont {Lagoyannis}, \citenamefont {L{\"o}nnroth}, \citenamefont
  {Mazurek}, \citenamefont {Norrby}, \citenamefont {Nyhus}, \citenamefont
  {Perdikakis}, \citenamefont {Schiller}, \citenamefont {Siem}, \citenamefont
  {Spyrou}, \citenamefont {Syed}, \citenamefont {Toft}, \citenamefont
  {Tveten},\ and\ \citenamefont {Voinov}}]{NuclFermiQuasi}%
  \BibitemOpen
  \bibfield  {author} {\bibinfo {author} {\bibfnamefont {M.}~\bibnamefont
  {Guttormsen}}, \bibinfo {author} {\bibfnamefont {A.~C.}\ \bibnamefont
  {Larsen}}, \bibinfo {author} {\bibfnamefont {A.}~\bibnamefont {B{\"u}rger}},
  \bibinfo {author} {\bibfnamefont {A.}~\bibnamefont {G{\"o}rgen}}, \bibinfo
  {author} {\bibfnamefont {S.}~\bibnamefont {Harissopulos}}, \bibinfo {author}
  {\bibfnamefont {M.}~\bibnamefont {Kmiecik}}, \bibinfo {author} {\bibfnamefont
  {T.}~\bibnamefont {Konstantinopoulos}}, \bibinfo {author} {\bibfnamefont
  {M.}~\bibnamefont {Krti{\v c}ka}}, \bibinfo {author} {\bibfnamefont
  {A.}~\bibnamefont {Lagoyannis}}, \bibinfo {author} {\bibfnamefont
  {T.}~\bibnamefont {L{\"o}nnroth}}, \bibinfo {author} {\bibfnamefont
  {K.}~\bibnamefont {Mazurek}}, \bibinfo {author} {\bibfnamefont
  {M.}~\bibnamefont {Norrby}}, \bibinfo {author} {\bibfnamefont {H.~T.}\
  \bibnamefont {Nyhus}}, \bibinfo {author} {\bibfnamefont {G.}~\bibnamefont
  {Perdikakis}}, \bibinfo {author} {\bibfnamefont {A.}~\bibnamefont
  {Schiller}}, \bibinfo {author} {\bibfnamefont {S.}~\bibnamefont {Siem}},
  \bibinfo {author} {\bibfnamefont {A.}~\bibnamefont {Spyrou}}, \bibinfo
  {author} {\bibfnamefont {N.~U.~H.}\ \bibnamefont {Syed}}, \bibinfo {author}
  {\bibfnamefont {H.~K.}\ \bibnamefont {Toft}}, \bibinfo {author}
  {\bibfnamefont {G.~M.}\ \bibnamefont {Tveten}}, \ and\ \bibinfo {author}
  {\bibfnamefont {A.}~\bibnamefont {Voinov}},\ }\bibfield  {title} {\enquote
  {\bibinfo {title} {Fermi's golden rule applied to the $\gamma$ decay in the
  quasicontinuum of $^{46}${T}i},}\ }\href
  {https://journals.aps.org/prc/abstract/10.1103/PhysRevC.83.014312} {\bibfield
   {journal} {\bibinfo  {journal} {Phys. Rev. C}\ }\textbf {\bibinfo {volume}
  {83}},\ \bibinfo {pages} {014312} (\bibinfo {year} {2011})}\BibitemShut
  {NoStop}%
\bibitem [{\citenamefont {Weinberg}(1995)}]{Weinberg1}%
  \BibitemOpen
  \bibfield  {author} {\bibinfo {author} {\bibfnamefont {S.}~\bibnamefont
  {Weinberg}},\ }\href@noop {} {\emph {\bibinfo {title} {The Quantum Theory of
  Fields}}},\ \bibinfo {edition} {1st}\ ed.,\ Vol.~\bibinfo {volume} {1}\
  (\bibinfo  {publisher} {Cambridge University Press},\ \bibinfo {year}
  {1995})\BibitemShut {NoStop}%
\bibitem [{\citenamefont {Schulze}\ and\ \citenamefont
  {Aichelin}(1989)}]{PartFermiFrag}%
  \BibitemOpen
  \bibfield  {author} {\bibinfo {author} {\bibfnamefont {H.~J.}\ \bibnamefont
  {Schulze}}\ and\ \bibinfo {author} {\bibfnamefont {J.}~\bibnamefont
  {Aichelin}},\ }\bibfield  {title} {\enquote {\bibinfo {title} {Does string
  fragmentation reveal more than longitudinal phase space?}}\ }\href
  {https://journals.aps.org/prd/abstract/10.1103/PhysRevD.39.3271} {\bibfield
  {journal} {\bibinfo  {journal} {Phys. Rev. D}\ }\textbf {\bibinfo {volume}
  {39}},\ \bibinfo {pages} {3271} (\bibinfo {year} {1989})}\BibitemShut
  {NoStop}%
\bibitem [{\citenamefont {Agarwal}\ and\ \citenamefont
  {Harshawardhan}(1996)}]{AtomFermiTwo}%
  \BibitemOpen
  \bibfield  {author} {\bibinfo {author} {\bibfnamefont {G.~S.}\ \bibnamefont
  {Agarwal}}\ and\ \bibinfo {author} {\bibfnamefont {W.}~\bibnamefont
  {Harshawardhan}},\ }\bibfield  {title} {\enquote {\bibinfo {title}
  {Inhibition and enhancement of two photon absorption},}\ }\href
  {https://journals.aps.org/prl/abstract/10.1103/PhysRevLett.77.1039}
  {\bibfield  {journal} {\bibinfo  {journal} {Phys. Rev. Lett.}\ }\textbf
  {\bibinfo {volume} {77}},\ \bibinfo {pages} {1039} (\bibinfo {year}
  {1996})}\BibitemShut {NoStop}%
\bibitem [{\citenamefont {Matloob}(2000)}]{AtomFermiMirror}%
  \BibitemOpen
  \bibfield  {author} {\bibinfo {author} {\bibfnamefont {R.}~\bibnamefont
  {Matloob}},\ }\bibfield  {title} {\enquote {\bibinfo {title} {Radiative
  properties of an atom in the vicinity of a mirror},}\ }\href
  {https://journals.aps.org/pra/abstract/10.1103/PhysRevA.62.022113} {\bibfield
   {journal} {\bibinfo  {journal} {Phys. Rev. A}\ }\textbf {\bibinfo {volume}
  {62}},\ \bibinfo {pages} {022113} (\bibinfo {year} {2000})}\BibitemShut
  {NoStop}%
\bibitem [{\citenamefont {Hettler}\ \emph {et~al.}(2003)\citenamefont
  {Hettler}, \citenamefont {Wenzel}, \citenamefont {Wegewijs},\ and\
  \citenamefont {Schoeller}}]{MolecFermiBenzene}%
  \BibitemOpen
  \bibfield  {author} {\bibinfo {author} {\bibfnamefont {M.~H.}\ \bibnamefont
  {Hettler}}, \bibinfo {author} {\bibfnamefont {W.}~\bibnamefont {Wenzel}},
  \bibinfo {author} {\bibfnamefont {M.~R.}\ \bibnamefont {Wegewijs}}, \ and\
  \bibinfo {author} {\bibfnamefont {H.}~\bibnamefont {Schoeller}},\ }\bibfield
  {title} {\enquote {\bibinfo {title} {Current collapse in tunneling transport
  through benzene},}\ }\href
  {https://journals.aps.org/prl/abstract/10.1103/PhysRevLett.90.076805}
  {\bibfield  {journal} {\bibinfo  {journal} {Phys. Rev. Lett.}\ }\textbf
  {\bibinfo {volume} {90}},\ \bibinfo {pages} {076805} (\bibinfo {year}
  {2003})}\BibitemShut {NoStop}%
\bibitem [{\citenamefont {Ghosh}\ \emph {et~al.}(2003)\citenamefont {Ghosh},
  \citenamefont {Bhattacharyya},\ and\ \citenamefont
  {Saha}}]{MolecFermiDissoc}%
  \BibitemOpen
  \bibfield  {author} {\bibinfo {author} {\bibfnamefont {S.}~\bibnamefont
  {Ghosh}}, \bibinfo {author} {\bibfnamefont {S.~S.}\ \bibnamefont
  {Bhattacharyya}}, \ and\ \bibinfo {author} {\bibfnamefont {S.}~\bibnamefont
  {Saha}},\ }\bibfield  {title} {\enquote {\bibinfo {title} {Reexamination of
  the photodissociation of {N}a{H}},}\ }\href
  {https://journals.aps.org/pra/abstract/10.1103/PhysRevA.67.054701} {\bibfield
   {journal} {\bibinfo  {journal} {Phys. Rev. A}\ }\textbf {\bibinfo {volume}
  {67}},\ \bibinfo {pages} {054701} (\bibinfo {year} {2003})}\BibitemShut
  {NoStop}%
\bibitem [{\citenamefont {Dharma-wardana}\ and\ \citenamefont
  {Perrot}(1998)}]{PlasmaFermiQuasi}%
  \BibitemOpen
  \bibfield  {author} {\bibinfo {author} {\bibfnamefont {M.~W.~C.}\
  \bibnamefont {Dharma-wardana}}\ and\ \bibinfo {author} {\bibfnamefont
  {F.}~\bibnamefont {Perrot}},\ }\bibfield  {title} {\enquote {\bibinfo {title}
  {Energy relaxation and the quasiequation of state of a dense two-temperature
  nonequilibrium plasma},}\ }\href
  {https://journals.aps.org/pre/abstract/10.1103/PhysRevE.58.3705} {\bibfield
  {journal} {\bibinfo  {journal} {Phys. Rev. E}\ }\textbf {\bibinfo {volume}
  {58}},\ \bibinfo {pages} {3705} (\bibinfo {year} {1998})}\BibitemShut
  {NoStop}%
\bibitem [{\citenamefont {Damjanovi{\v c}}\ \emph {et~al.}(1999)\citenamefont
  {Damjanovi{\v c}}, \citenamefont {Ritz},\ and\ \citenamefont
  {Schulten}}]{BioFermiPurple}%
  \BibitemOpen
  \bibfield  {author} {\bibinfo {author} {\bibfnamefont {A.}~\bibnamefont
  {Damjanovi{\v c}}}, \bibinfo {author} {\bibfnamefont {T.}~\bibnamefont
  {Ritz}}, \ and\ \bibinfo {author} {\bibfnamefont {K.}~\bibnamefont
  {Schulten}},\ }\bibfield  {title} {\enquote {\bibinfo {title} {Energy
  transfer between carotenoids and bacteriochlorophylls in light-harvesting
  complex {\Rmnum{2}} of purple bacteria},}\ }\href
  {https://journals.aps.org/pre/abstract/10.1103/PhysRevE.59.3293} {\bibfield
  {journal} {\bibinfo  {journal} {Phys. Rev. E}\ }\textbf {\bibinfo {volume}
  {59}},\ \bibinfo {pages} {3293} (\bibinfo {year} {1999})}\BibitemShut
  {NoStop}%
\bibitem [{\citenamefont {Hodges}\ \emph {et~al.}(1971)\citenamefont {Hodges},
  \citenamefont {Smith},\ and\ \citenamefont {Wilkins}}]{SolidFermiSurface}%
  \BibitemOpen
  \bibfield  {author} {\bibinfo {author} {\bibfnamefont {C.}~\bibnamefont
  {Hodges}}, \bibinfo {author} {\bibfnamefont {H.}~\bibnamefont {Smith}}, \
  and\ \bibinfo {author} {\bibfnamefont {J.~W.}\ \bibnamefont {Wilkins}},\
  }\bibfield  {title} {\enquote {\bibinfo {title} {Effect of {F}ermi surface
  geometry on electron-electron scattering},}\ }\href
  {https://journals.aps.org/prb/abstract/10.1103/PhysRevB.4.302} {\bibfield
  {journal} {\bibinfo  {journal} {Phys. Rev. B}\ }\textbf {\bibinfo {volume}
  {4}},\ \bibinfo {pages} {302} (\bibinfo {year} {1971})}\BibitemShut {NoStop}%
\bibitem [{\citenamefont {Ray}\ and\ \citenamefont
  {Basur}(1992)}]{SolidFermiExciton}%
  \BibitemOpen
  \bibfield  {author} {\bibinfo {author} {\bibfnamefont {P.}~\bibnamefont
  {Ray}}\ and\ \bibinfo {author} {\bibfnamefont {P.~K.}\ \bibnamefont
  {Basur}},\ }\bibfield  {title} {\enquote {\bibinfo {title} {Low-temperature
  exciton linewidth in short-period superlattices},}\ }\href
  {https://journals.aps.org/prb/abstract/10.1103/PhysRevB.46.13268} {\bibfield
  {journal} {\bibinfo  {journal} {Phys. Rev. B}\ }\textbf {\bibinfo {volume}
  {46}},\ \bibinfo {pages} {13268} (\bibinfo {year} {1992})}\BibitemShut
  {NoStop}%
\bibitem [{\citenamefont {Kenkre}\ and\ \citenamefont
  {Parris}(2002)}]{SolidFermiOrgCry}%
  \BibitemOpen
  \bibfield  {author} {\bibinfo {author} {\bibfnamefont {V.~M.}\ \bibnamefont
  {Kenkre}}\ and\ \bibinfo {author} {\bibfnamefont {P.~E.}\ \bibnamefont
  {Parris}},\ }\bibfield  {title} {\enquote {\bibinfo {title} {Mechanism for
  carrier velocity saturation in pure organic crystals},}\ }\href
  {https://journals.aps.org/prb/abstract/10.1103/PhysRevB.65.245106} {\bibfield
   {journal} {\bibinfo  {journal} {Phys. Rev. B}\ }\textbf {\bibinfo {volume}
  {65}},\ \bibinfo {pages} {245106} (\bibinfo {year} {2002})}\BibitemShut
  {NoStop}%
\bibitem [{\citenamefont {Hoeppe}\ \emph {et~al.}(2012)\citenamefont {Hoeppe},
  \citenamefont {Wolff}, \citenamefont {K{\"u}chenmeister}, \citenamefont
  {Niegemann}, \citenamefont {Drescher}, \citenamefont {Benner},\ and\
  \citenamefont {Busch}}]{PhotoFermiDirect}%
  \BibitemOpen
  \bibfield  {author} {\bibinfo {author} {\bibfnamefont {U.}~\bibnamefont
  {Hoeppe}}, \bibinfo {author} {\bibfnamefont {C.}~\bibnamefont {Wolff}},
  \bibinfo {author} {\bibfnamefont {J.}~\bibnamefont {K{\"u}chenmeister}},
  \bibinfo {author} {\bibfnamefont {J.}~\bibnamefont {Niegemann}}, \bibinfo
  {author} {\bibfnamefont {M.}~\bibnamefont {Drescher}}, \bibinfo {author}
  {\bibfnamefont {H.}~\bibnamefont {Benner}}, \ and\ \bibinfo {author}
  {\bibfnamefont {K.}~\bibnamefont {Busch}},\ }\bibfield  {title} {\enquote
  {\bibinfo {title} {Direct observation of non-markovian radiation dynamics in
  3{D} bulk photonic crystals},}\ }\href
  {https://journals.aps.org/prl/abstract/10.1103/PhysRevLett.108.043603}
  {\bibfield  {journal} {\bibinfo  {journal} {Phys. Rev. Lett.}\ }\textbf
  {\bibinfo {volume} {108}},\ \bibinfo {pages} {043603} (\bibinfo {year}
  {2012})}\BibitemShut {NoStop}%
\bibitem [{\citenamefont {Weisskopf}\ and\ \citenamefont {Wigner}(1930)}]{WW}%
  \BibitemOpen
  \bibfield  {author} {\bibinfo {author} {\bibfnamefont {V.}~\bibnamefont
  {Weisskopf}}\ and\ \bibinfo {author} {\bibfnamefont {E.~P.}\ \bibnamefont
  {Wigner}},\ }\bibfield  {title} {\enquote {\bibinfo {title} {Berechnung der
  nat{\"u}rlichen {L}inienbreite auf {G}rund der {D}iracschen
  {L}ichttheorie},}\ }\href
  {http://link.springer.com/article/10.1007%2FBF01336768} {\bibfield  {journal}
  {\bibinfo  {journal} {Z. Phys.}\ }\textbf {\bibinfo {volume} {63}},\ \bibinfo
  {pages} {54} (\bibinfo {year} {1930})}\BibitemShut {NoStop}%
\bibitem [{\citenamefont {Khalfin}(1968)}]{Khalfin}%
  \BibitemOpen
  \bibfield  {author} {\bibinfo {author} {\bibfnamefont {L.~A.}\ \bibnamefont
  {Khalfin}},\ }\bibfield  {title} {\enquote {\bibinfo {title}
  {Phenomenological theory of {K}$^0$ mesons and the non-exponential character
  of the decay},}\ }\href@noop {} {\bibfield  {journal} {\bibinfo  {journal}
  {JETP Lett.}\ }\textbf {\bibinfo {volume} {8}},\ \bibinfo {pages} {106}
  (\bibinfo {year} {1968})}\BibitemShut {NoStop}%
\bibitem [{\citenamefont {Facchi}(2000)}]{FacchiPhD}%
  \BibitemOpen
  \bibfield  {author} {\bibinfo {author} {\bibfnamefont {P.}~\bibnamefont
  {Facchi}},\ }\emph {\bibinfo {title} {Quantum Time Evolution: Free and
  Controlled Dynamics}},\ \href
  {http://www.ba.infn.it/~facchi/lectures/thesis.pdf} {\bibinfo {type} {Ph{D}
  {T}hesis}},\ \bibinfo  {school} {UniversitÃ degli Studi di Bari} (\bibinfo
  {year} {2000})\BibitemShut {NoStop}%
\bibitem [{\citenamefont {Misra}\ and\ \citenamefont
  {Sudarshan}(1977)}]{MisraSudarshan}%
  \BibitemOpen
  \bibfield  {author} {\bibinfo {author} {\bibfnamefont {B.}~\bibnamefont
  {Misra}}\ and\ \bibinfo {author} {\bibfnamefont {E.~C.~G.}\ \bibnamefont
  {Sudarshan}},\ }\bibfield  {title} {\enquote {\bibinfo {title} {The {Z}eno's
  paradox in quantum theory},}\ }\href
  {http://scitation.aip.org/content/aip/journal/jmp/18/4/10.1063/1.523304}
  {\bibfield  {journal} {\bibinfo  {journal} {J. Math. Phys.}\ }\textbf
  {\bibinfo {volume} {18}},\ \bibinfo {pages} {756} (\bibinfo {year}
  {1977})}\BibitemShut {NoStop}%
\bibitem [{\citenamefont {Kofman}\ and\ \citenamefont
  {Kurizki}(2000)}]{KofKur}%
  \BibitemOpen
  \bibfield  {author} {\bibinfo {author} {\bibfnamefont {A.~G.}\ \bibnamefont
  {Kofman}}\ and\ \bibinfo {author} {\bibfnamefont {G.}~\bibnamefont
  {Kurizki}},\ }\bibfield  {title} {\enquote {\bibinfo {title} {Acceleration of
  quantum decay processes by frequent observations},}\ }\href
  {https://www.nature.com/articles/35014537} {\bibfield  {journal} {\bibinfo
  {journal} {Nature}\ }\textbf {\bibinfo {volume} {405}},\ \bibinfo {pages}
  {546} (\bibinfo {year} {2000})}\BibitemShut {NoStop}%
\bibitem [{\citenamefont {Facchi}\ \emph {et~al.}(2001)\citenamefont {Facchi},
  \citenamefont {Nakazato},\ and\ \citenamefont {Pascazio}}]{FaNaPa}%
  \BibitemOpen
  \bibfield  {author} {\bibinfo {author} {\bibfnamefont {P.}~\bibnamefont
  {Facchi}}, \bibinfo {author} {\bibfnamefont {H.}~\bibnamefont {Nakazato}}, \
  and\ \bibinfo {author} {\bibfnamefont {S.}~\bibnamefont {Pascazio}},\
  }\bibfield  {title} {\enquote {\bibinfo {title} {From the quantum {Z}eno to
  the inverse quantum {Z}eno effect},}\ }\href
  {https://journals.aps.org/prl/abstract/10.1103/PhysRevLett.86.2699}
  {\bibfield  {journal} {\bibinfo  {journal} {Phys. Rev. Lett.}\ }\textbf
  {\bibinfo {volume} {86}},\ \bibinfo {pages} {2699} (\bibinfo {year}
  {2001})}\BibitemShut {NoStop}%
\bibitem [{\citenamefont {Antoniou}\ \emph {et~al.}(2001)\citenamefont
  {Antoniou}, \citenamefont {Karpov}, \citenamefont {Pronko},\ and\
  \citenamefont {Yarevsky}}]{KarpovZeno}%
  \BibitemOpen
  \bibfield  {author} {\bibinfo {author} {\bibfnamefont {I.}~\bibnamefont
  {Antoniou}}, \bibinfo {author} {\bibfnamefont {E.}~\bibnamefont {Karpov}},
  \bibinfo {author} {\bibfnamefont {G.}~\bibnamefont {Pronko}}, \ and\ \bibinfo
  {author} {\bibfnamefont {E.}~\bibnamefont {Yarevsky}},\ }\bibfield  {title}
  {\enquote {\bibinfo {title} {Quantum {Z}eno and anti-{Z}eno effects in the
  {F}riedrichs model},}\ }\href
  {https://journals.aps.org/pra/abstract/10.1103/PhysRevA.63.062110} {\bibfield
   {journal} {\bibinfo  {journal} {Phys. Rev. A}\ }\textbf {\bibinfo {volume}
  {63}},\ \bibinfo {pages} {062110} (\bibinfo {year} {2001})}\BibitemShut
  {NoStop}%
\bibitem [{\citenamefont {Bellac}(2013)}]{Bellac}%
  \BibitemOpen
  \bibfield  {author} {\bibinfo {author} {\bibfnamefont {M.~Le}\ \bibnamefont
  {Bellac}},\ }\href@noop {} {\emph {\bibinfo {title} {Quantum Physics}}}\
  (\bibinfo  {publisher} {Cambridge University Press},\ \bibinfo {year}
  {2013})\BibitemShut {NoStop}%
\bibitem [{\citenamefont {Englert}(2006)}]{Englert}%
  \BibitemOpen
  \bibfield  {author} {\bibinfo {author} {\bibfnamefont {B.~G.}\ \bibnamefont
  {Englert}},\ }\href@noop {} {\emph {\bibinfo {title} {Lectures On Quantum
  Mechanics}}},\ Vol.~\bibinfo {volume} {3}\ (\bibinfo  {publisher} {World
  Scientific},\ \bibinfo {year} {2006})\BibitemShut {NoStop}%
\bibitem [{\citenamefont {Debierre}\ \emph {et~al.}(2015)\citenamefont
  {Debierre}, \citenamefont {Goessens}, \citenamefont {Brainis},\ and\
  \citenamefont {Durt}}]{EdouardIsa}%
  \BibitemOpen
  \bibfield  {author} {\bibinfo {author} {\bibfnamefont {V.}~\bibnamefont
  {Debierre}}, \bibinfo {author} {\bibfnamefont {I.}~\bibnamefont {Goessens}},
  \bibinfo {author} {\bibfnamefont {\'{E}}\ \bibnamefont {Brainis}}, \ and\
  \bibinfo {author} {\bibfnamefont {T.}~\bibnamefont {Durt}},\ }\bibfield
  {title} {\enquote {\bibinfo {title} {Fermi golden rule beyond the {Z}eno
  regime},}\ }\href
  {http://journals.aps.org/pra/abstract/10.1103/PhysRevA.92.023825} {\bibfield
  {journal} {\bibinfo  {journal} {Phys. Rev. A}\ }\textbf {\bibinfo {volume}
  {92}},\ \bibinfo {pages} {023825} (\bibinfo {year} {2015})}\BibitemShut
  {NoStop}%
\bibitem [{\citenamefont {Messiah}(1965)}]{Messiah2}%
  \BibitemOpen
  \bibfield  {author} {\bibinfo {author} {\bibfnamefont {A.}~\bibnamefont
  {Messiah}},\ }\href@noop {} {\emph {\bibinfo {title} {M{\'e}canique
  Quantique}}},\ \bibinfo {edition} {1st}\ ed.,\ Vol.~\bibinfo {volume} {2}\
  (\bibinfo  {publisher} {Dunod},\ \bibinfo {year} {1965})\BibitemShut
  {NoStop}%
\bibitem [{\citenamefont {Cohen-Tannoudji}\ \emph {et~al.}(1996)\citenamefont
  {Cohen-Tannoudji}, \citenamefont {Diu},\ and\ \citenamefont
  {Lalo{\"e}}}]{Cohen2}%
  \BibitemOpen
  \bibfield  {author} {\bibinfo {author} {\bibfnamefont {C.}~\bibnamefont
  {Cohen-Tannoudji}}, \bibinfo {author} {\bibfnamefont {B.}~\bibnamefont
  {Diu}}, \ and\ \bibinfo {author} {\bibfnamefont {F.}~\bibnamefont
  {Lalo{\"e}}},\ }\href@noop {} {\emph {\bibinfo {title} {M{\'e}canique
  Quantique}}},\ \bibinfo {edition} {2nd}\ ed.,\ Vol.~\bibinfo {volume} {2}\
  (\bibinfo  {publisher} {Hermann},\ \bibinfo {year} {1996})\BibitemShut
  {NoStop}%
\bibitem [{\citenamefont {Grynberg}\ \emph {et~al.}(2010)\citenamefont
  {Grynberg}, \citenamefont {Aspect},\ and\ \citenamefont
  {Fabre}}]{grynberg2010introduction}%
  \BibitemOpen
  \bibfield  {author} {\bibinfo {author} {\bibfnamefont {G.}~\bibnamefont
  {Grynberg}}, \bibinfo {author} {\bibfnamefont {A.}~\bibnamefont {Aspect}}, \
  and\ \bibinfo {author} {\bibfnamefont {C.}~\bibnamefont {Fabre}},\
  }\href@noop {} {\emph {\bibinfo {title} {Introduction to quantum optics: from
  the semi-classical approach to quantized light}}}\ (\bibinfo  {publisher}
  {Cambridge university press},\ \bibinfo {year} {2010})\BibitemShut {NoStop}%
\bibitem [{\citenamefont {Barnett}\ and\ \citenamefont
  {Radmore}(2002)}]{barnett2002methods}%
  \BibitemOpen
  \bibfield  {author} {\bibinfo {author} {\bibfnamefont {S.}~\bibnamefont
  {Barnett}}\ and\ \bibinfo {author} {\bibfnamefont {P.~M.}\ \bibnamefont
  {Radmore}},\ }\href@noop {} {\emph {\bibinfo {title} {Methods in theoretical
  quantum optics}}},\ Vol.~\bibinfo {volume} {15}\ (\bibinfo  {publisher}
  {Oxford University Press},\ \bibinfo {year} {2002})\BibitemShut {NoStop}%
\bibitem [{\citenamefont {Piraux}\ \emph {et~al.}(1990)\citenamefont {Piraux},
  \citenamefont {Bhatt},\ and\ \citenamefont {Knight}}]{Threshold}%
  \BibitemOpen
  \bibfield  {author} {\bibinfo {author} {\bibfnamefont {B.}~\bibnamefont
  {Piraux}}, \bibinfo {author} {\bibfnamefont {R.}~\bibnamefont {Bhatt}}, \
  and\ \bibinfo {author} {\bibfnamefont {P.~L.}\ \bibnamefont {Knight}},\
  }\bibfield  {title} {\enquote {\bibinfo {title} {Near-threshold excitation of
  continuum resonances},}\ }\href
  {https://journals.aps.org/pra/abstract/10.1103/PhysRevA.41.6296} {\bibfield
  {journal} {\bibinfo  {journal} {Phys. Rev. A}\ }\textbf {\bibinfo {volume}
  {41}},\ \bibinfo {pages} {6296} (\bibinfo {year} {1990})}\BibitemShut
  {NoStop}%
\bibitem [{\citenamefont {Seke}(1994)}]{Seke}%
  \BibitemOpen
  \bibfield  {author} {\bibinfo {author} {\bibfnamefont {J.}~\bibnamefont
  {Seke}},\ }\bibfield  {title} {\enquote {\bibinfo {title} {Analytic
  evaluation of exact transition matrix elements in nonrelativistic hydrogenic
  atoms},}\ }\href
  {http://www.sciencedirect.com/science/article/pii/0378437194901562}
  {\bibfield  {journal} {\bibinfo  {journal} {Phys. A}\ }\textbf {\bibinfo
  {volume} {203}},\ \bibinfo {pages} {269} (\bibinfo {year}
  {1994})}\BibitemShut {NoStop}%
\bibitem [{\citenamefont {Lassalle}\ \emph {et~al.}(2018)\citenamefont
  {Lassalle}, \citenamefont {Champenois}, \citenamefont {Stout}, \citenamefont
  {Debierre},\ and\ \citenamefont {Durt}}]{LassalleChampenois}%
  \BibitemOpen
  \bibfield  {author} {\bibinfo {author} {\bibfnamefont {E.}~\bibnamefont
  {Lassalle}}, \bibinfo {author} {\bibfnamefont {C.}~\bibnamefont
  {Champenois}}, \bibinfo {author} {\bibfnamefont {B.}~\bibnamefont {Stout}},
  \bibinfo {author} {\bibfnamefont {V.}~\bibnamefont {Debierre}}, \ and\
  \bibinfo {author} {\bibfnamefont {T.}~\bibnamefont {Durt}},\ }\bibfield
  {title} {\enquote {\bibinfo {title} {Conditions for anti-{Z}eno-effect
  observation in free-space atomic radiative decay},}\ }\href
  {https://journals.aps.org/pra/abstract/10.1103/PhysRevA.97.062122} {\bibfield
   {journal} {\bibinfo  {journal} {Phys. Rev. A}\ }\textbf {\bibinfo {volume}
  {97}},\ \bibinfo {pages} {062122} (\bibinfo {year} {2018})}\BibitemShut
  {NoStop}%
\bibitem [{\citenamefont {Lassalle}(2019)}]{lassalle:tel-02283698}%
  \BibitemOpen
  \bibfield  {author} {\bibinfo {author} {\bibfnamefont {Emmanuel}\
  \bibnamefont {Lassalle}},\ }\emph {\bibinfo {title} {{Environment induced
  modifications of spontaneous quantum emission}}},\ \href
  {https://tel.archives-ouvertes.fr/tel-02283698} {\bibinfo {type} {{P}h{D}
  {T}hesis}},\ \bibinfo  {school} {{Aix Marseille Universit{\'e}}} (\bibinfo
  {year} {2019})\BibitemShut {NoStop}%
\bibitem [{\citenamefont {Zhang}\ \emph {et~al.}(2018)\citenamefont {Zhang},
  \citenamefont {Jing}, \citenamefont {Wang},\ and\ \citenamefont
  {Zhu}}]{ChinaZenoCriterion}%
  \BibitemOpen
  \bibfield  {author} {\bibinfo {author} {\bibfnamefont {J.~M.}\ \bibnamefont
  {Zhang}}, \bibinfo {author} {\bibfnamefont {J.}~\bibnamefont {Jing}},
  \bibinfo {author} {\bibfnamefont {L.~G.}\ \bibnamefont {Wang}}, \ and\
  \bibinfo {author} {\bibfnamefont {S.~Y.}\ \bibnamefont {Zhu}},\ }\bibfield
  {title} {\enquote {\bibinfo {title} {Criterion for quantum {Z}eno and
  anti-{Z}eno effects},}\ }\href
  {https://journals.aps.org/pra/abstract/10.1103/PhysRevA.98.012135} {\bibfield
   {journal} {\bibinfo  {journal} {Phys. Rev. A}\ }\textbf {\bibinfo {volume}
  {98}},\ \bibinfo {pages} {012135} (\bibinfo {year} {2018})}\BibitemShut
  {NoStop}%
\bibitem [{\citenamefont {Leggett}\ \emph {et~al.}(1987)\citenamefont
  {Leggett}, \citenamefont {Chakravarty}, \citenamefont {Dorsey}, \citenamefont
  {Fisher}, \citenamefont {Garg},\ and\ \citenamefont {Zwerger}}]{OhmBaths}%
  \BibitemOpen
  \bibfield  {author} {\bibinfo {author} {\bibfnamefont {A.~J.}\ \bibnamefont
  {Leggett}}, \bibinfo {author} {\bibfnamefont {S.}~\bibnamefont
  {Chakravarty}}, \bibinfo {author} {\bibfnamefont {A.~T.}\ \bibnamefont
  {Dorsey}}, \bibinfo {author} {\bibfnamefont {M.~P.~A.}\ \bibnamefont
  {Fisher}}, \bibinfo {author} {\bibfnamefont {A.}~\bibnamefont {Garg}}, \ and\
  \bibinfo {author} {\bibfnamefont {W.}~\bibnamefont {Zwerger}},\ }\bibfield
  {title} {\enquote {\bibinfo {title} {Dynamics of the dissipative two-state
  system},}\ }\href
  {http://journals.aps.org/rmp/abstract/10.1103/RevModPhys.59.1} {\bibfield
  {journal} {\bibinfo  {journal} {Rev. Mod. Phys.}\ }\textbf {\bibinfo {volume}
  {59}},\ \bibinfo {pages} {1} (\bibinfo {year} {1987})}\BibitemShut {NoStop}%
\bibitem [{\citenamefont {Zawadowski}\ and\ \citenamefont
  {Zim\'{a}nyi}(1985)}]{TheoryTunnelInteract}%
  \BibitemOpen
  \bibfield  {author} {\bibinfo {author} {\bibfnamefont {A.}~\bibnamefont
  {Zawadowski}}\ and\ \bibinfo {author} {\bibfnamefont {G.~T.}\ \bibnamefont
  {Zim\'{a}nyi}},\ }\bibfield  {title} {\enquote {\bibinfo {title} {Theory of
  tunneling of an atom interacting with a degenerate electron gas},}\ }\href
  {https://journals.aps.org/prb/abstract/10.1103/PhysRevB.32.1373} {\bibfield
  {journal} {\bibinfo  {journal} {Phys. Rev. B}\ }\textbf {\bibinfo {volume}
  {21}},\ \bibinfo {pages} {1373(R)} (\bibinfo {year} {1985})}\BibitemShut
  {NoStop}%
\bibitem [{\citenamefont {Muramatsu}\ and\ \citenamefont
  {Guinea}(1986)}]{LowTTheoryTunnelInteract}%
  \BibitemOpen
  \bibfield  {author} {\bibinfo {author} {\bibfnamefont {A.}~\bibnamefont
  {Muramatsu}}\ and\ \bibinfo {author} {\bibfnamefont {F.}~\bibnamefont
  {Guinea}},\ }\bibfield  {title} {\enquote {\bibinfo {title} {Low-temperature
  behavior of a tunneling atom interacting with a degenerate electron gas},}\
  }\href {https://journals.aps.org/prl/abstract/10.1103/PhysRevLett.57.2337}
  {\bibfield  {journal} {\bibinfo  {journal} {Phys. Rev. Lett.}\ }\textbf
  {\bibinfo {volume} {57}},\ \bibinfo {pages} {2337} (\bibinfo {year}
  {1986})}\BibitemShut {NoStop}%
\bibitem [{Note1()}]{Note1}%
  \BibitemOpen
  \bibinfo {note} {For instance, for atomic transitions, $F_{\protect \mathrm
  {X}}\left (\omega \right )=\left [1+\left (\omega /\omega _{\protect \mathrm
  {X}}\right )^2\right ]^{-\mu }$ with $\mu \geq 4$ an integer, and for sub- or
  super-Ohmic baths, $F_{\protect \mathrm {X}}\left (\omega \right )=\protect
  \qopname \relax o{exp}\left (-\omega /\omega _{\protect \mathrm {X}}\right
  )$.}\BibitemShut {Stop}%
\bibitem [{\citenamefont {Moses}(1973)}]{Moses}%
  \BibitemOpen
  \bibfield  {author} {\bibinfo {author} {\bibfnamefont {H.~E.}\ \bibnamefont
  {Moses}},\ }\bibfield  {title} {\enquote {\bibinfo {title} {Photon wave
  functions and the exact electromagnetic matrix elements for hydrogenic
  atoms},}\ }\href
  {https://journals.aps.org/pra/abstract/10.1103/PhysRevA.8.1710} {\bibfield
  {journal} {\bibinfo  {journal} {Phys. Rev. A}\ }\textbf {\bibinfo {volume}
  {8}},\ \bibinfo {pages} {1710} (\bibinfo {year} {1973})}\BibitemShut
  {NoStop}%
\bibitem [{Note2()}]{Note2}%
  \BibitemOpen
  \bibinfo {note} {We call $\omega _0t\gg 1$ the resonant regime because in
  this regime, the transition frequency $\omega _0$ of the two-level system has
  been spectrally resolved by the dynamics and starts playing its role as the
  resonance frequency.}\BibitemShut {Stop}%
\bibitem [{Note3()}]{Note3}%
  \BibitemOpen
  \bibinfo {note} {Note also from Fig.~\ref {fig:BroadAll} that for $\eta \ll
  1$, the generalised decay rate verifies $\Gamma \left (t\right )\simeq \Gamma
  _0$ long before $t\sim 1/\omega _0$ but however, $\Gamma \left (t\right )$ is
  not constant even at the lowest order of approximation in this regime, which
  precludes from including it in the Fermi regime.}\BibitemShut {Stop}%
\bibitem [{\citenamefont {Raimond}\ and\ \citenamefont
  {Haroche}(2006)}]{raimond2006exploring}%
  \BibitemOpen
  \bibfield  {author} {\bibinfo {author} {\bibfnamefont {J.~M.}\ \bibnamefont
  {Raimond}}\ and\ \bibinfo {author} {\bibfnamefont {S.}~\bibnamefont
  {Haroche}},\ }\href@noop {} {\emph {\bibinfo {title} {Exploring the
  quantum}}}\ (\bibinfo  {publisher} {Oxford University Press, Oxford},\
  \bibinfo {year} {2006})\BibitemShut {NoStop}%
\bibitem [{\citenamefont {Ching}\ \emph {et~al.}(1987)\citenamefont {Ching},
  \citenamefont {Lai},\ and\ \citenamefont {Young}}]{ching1987dielectric}%
  \BibitemOpen
  \bibfield  {author} {\bibinfo {author} {\bibfnamefont {S.~C.}\ \bibnamefont
  {Ching}}, \bibinfo {author} {\bibfnamefont {H.~M.}\ \bibnamefont {Lai}}, \
  and\ \bibinfo {author} {\bibfnamefont {K.}~\bibnamefont {Young}},\ }\bibfield
   {title} {\enquote {\bibinfo {title} {Dielectric microspheres as optical
  cavities: Einstein {$A$} and {$B$} coefficients and level shift},}\ }\href
  {https://www.osapublishing.org/josab/abstract.cfm?uri=josab-4-12-2004}
  {\bibfield  {journal} {\bibinfo  {journal} {J. Opt. Soc. Am. B}\ }\textbf
  {\bibinfo {volume} {4}},\ \bibinfo {pages} {2004--2009} (\bibinfo {year}
  {1987})}\BibitemShut {NoStop}%
\bibitem [{\citenamefont {Dung}\ \emph {et~al.}(2001)\citenamefont {Dung},
  \citenamefont {Kn{\"o}ll},\ and\ \citenamefont {Welsch}}]{dung2001decay}%
  \BibitemOpen
  \bibfield  {author} {\bibinfo {author} {\bibfnamefont {H.~T.}\ \bibnamefont
  {Dung}}, \bibinfo {author} {\bibfnamefont {L.}~\bibnamefont {Kn{\"o}ll}}, \
  and\ \bibinfo {author} {\bibfnamefont {D.~G.}\ \bibnamefont {Welsch}},\
  }\bibfield  {title} {\enquote {\bibinfo {title} {Decay of an excited atom
  near an absorbing microsphere},}\ }\href
  {https://journals.aps.org/pra/abstract/10.1103/PhysRevA.64.013804} {\bibfield
   {journal} {\bibinfo  {journal} {Phys. Rev. A}\ }\textbf {\bibinfo {volume}
  {64}},\ \bibinfo {pages} {013804} (\bibinfo {year} {2001})}\BibitemShut
  {NoStop}%
\bibitem [{\citenamefont {Delga}\ \emph {et~al.}(2014)\citenamefont {Delga},
  \citenamefont {Feist}, \citenamefont {Bravo-Abad},\ and\ \citenamefont
  {Garcia-Vidal}}]{delga2014quantum}%
  \BibitemOpen
  \bibfield  {author} {\bibinfo {author} {\bibfnamefont {A.}~\bibnamefont
  {Delga}}, \bibinfo {author} {\bibfnamefont {J.}~\bibnamefont {Feist}},
  \bibinfo {author} {\bibfnamefont {J.}~\bibnamefont {Bravo-Abad}}, \ and\
  \bibinfo {author} {\bibfnamefont {F.~J.}\ \bibnamefont {Garcia-Vidal}},\
  }\bibfield  {title} {\enquote {\bibinfo {title} {Quantum emitters near a
  metal nanoparticle: strong coupling and quenching},}\ }\href
  {https://journals.aps.org/prl/abstract/10.1103/PhysRevLett.112.253601}
  {\bibfield  {journal} {\bibinfo  {journal} {Phys. Rev. Lett.}\ }\textbf
  {\bibinfo {volume} {112}},\ \bibinfo {pages} {253601} (\bibinfo {year}
  {2014})}\BibitemShut {NoStop}%
\bibitem [{\citenamefont {Varguet}\ \emph {et~al.}(2016)\citenamefont
  {Varguet}, \citenamefont {Rousseaux}, \citenamefont {Dzsotjan}, \citenamefont
  {Jauslin}, \citenamefont {Gu{\'e}rin},\ and\ \citenamefont {Colas~des
  Francs}}]{varguet2016dressed}%
  \BibitemOpen
  \bibfield  {author} {\bibinfo {author} {\bibfnamefont {H.}~\bibnamefont
  {Varguet}}, \bibinfo {author} {\bibfnamefont {B.}~\bibnamefont {Rousseaux}},
  \bibinfo {author} {\bibfnamefont {D.}~\bibnamefont {Dzsotjan}}, \bibinfo
  {author} {\bibfnamefont {H.~R.}\ \bibnamefont {Jauslin}}, \bibinfo {author}
  {\bibfnamefont {S.}~\bibnamefont {Gu{\'e}rin}}, \ and\ \bibinfo {author}
  {\bibfnamefont {G.}~\bibnamefont {Colas~des Francs}},\ }\bibfield  {title}
  {\enquote {\bibinfo {title} {Dressed states of a quantum emitter strongly
  coupled to a metal nanoparticle},}\ }\href
  {https://www.osapublishing.org/ol/abstract.cfm?uri=ol-41-19-4480} {\bibfield
  {journal} {\bibinfo  {journal} {Opt. Lett.}\ }\textbf {\bibinfo {volume}
  {41}},\ \bibinfo {pages} {4480--4483} (\bibinfo {year} {2016})}\BibitemShut
  {NoStop}%
\bibitem [{\citenamefont {Aljunid}\ \emph {et~al.}(2016)\citenamefont
  {Aljunid}, \citenamefont {Chan}, \citenamefont {Adamo}, \citenamefont
  {Ducloy}, \citenamefont {Wilkowski},\ and\ \citenamefont
  {Zheludev}}]{aljunid2016atomic}%
  \BibitemOpen
  \bibfield  {author} {\bibinfo {author} {\bibfnamefont {S.~A.}\ \bibnamefont
  {Aljunid}}, \bibinfo {author} {\bibfnamefont {E.~A.}\ \bibnamefont {Chan}},
  \bibinfo {author} {\bibfnamefont {G.}~\bibnamefont {Adamo}}, \bibinfo
  {author} {\bibfnamefont {M.}~\bibnamefont {Ducloy}}, \bibinfo {author}
  {\bibfnamefont {D.}~\bibnamefont {Wilkowski}}, \ and\ \bibinfo {author}
  {\bibfnamefont {N.~I.}\ \bibnamefont {Zheludev}},\ }\bibfield  {title}
  {\enquote {\bibinfo {title} {Atomic response in the near-field of
  nanostructured plasmonic metamaterial},}\ }\href
  {https://pubs.acs.org/doi/abs/10.1021/acs.nanolett.6b00446} {\bibfield
  {journal} {\bibinfo  {journal} {Nano Lett.}\ }\textbf {\bibinfo {volume}
  {16}},\ \bibinfo {pages} {3137--3141} (\bibinfo {year} {2016})}\BibitemShut
  {NoStop}%
\bibitem [{\citenamefont {Jiang}\ \emph {et~al.}(2017)\citenamefont {Jiang},
  \citenamefont {Qi}, \citenamefont {Zhang}, \citenamefont {Sun}, \citenamefont
  {Chen}, \citenamefont {Chen}, \citenamefont {Yu}, \citenamefont {Li},\ and\
  \citenamefont {Tian}}]{jiang2017ultra}%
  \BibitemOpen
  \bibfield  {author} {\bibinfo {author} {\bibfnamefont {M.}~\bibnamefont
  {Jiang}}, \bibinfo {author} {\bibfnamefont {J.}~\bibnamefont {Qi}}, \bibinfo
  {author} {\bibfnamefont {M.}~\bibnamefont {Zhang}}, \bibinfo {author}
  {\bibfnamefont {Q.}~\bibnamefont {Sun}}, \bibinfo {author} {\bibfnamefont
  {J.}~\bibnamefont {Chen}}, \bibinfo {author} {\bibfnamefont {Z.}~\bibnamefont
  {Chen}}, \bibinfo {author} {\bibfnamefont {X.}~\bibnamefont {Yu}}, \bibinfo
  {author} {\bibfnamefont {Y.}~\bibnamefont {Li}}, \ and\ \bibinfo {author}
  {\bibfnamefont {J.}~\bibnamefont {Tian}},\ }\bibfield  {title} {\enquote
  {\bibinfo {title} {Ultra-high quality factor metallic micro-cavity based on
  concentric double metal-insulator-metal rings},}\ }\href
  {https://www.nature.com/articles/s41598-017-15906-4} {\bibfield  {journal}
  {\bibinfo  {journal} {Sci. Rep.}\ }\textbf {\bibinfo {volume} {7}},\ \bibinfo
  {pages} {15663} (\bibinfo {year} {2017})}\BibitemShut {NoStop}%
\end{thebibliography}%
\end{document}